\documentclass[10pt,pra,letterpaper,typeset,superscriptaddress,twocolumn,showpacs,spanish,english,floatfix]{revtex4-1}

\usepackage{color}
\usepackage{graphicx}
\usepackage{amsmath}
\usepackage{amssymb,amsthm}
\usepackage{bm}

\usepackage{hyperref}
\hypersetup{
pdftitle={Polarization dependent coupling},
anchorcolor=blue,
colorlinks=true,
citecolor=blue,
urlcolor=blue,
linkcolor=blue}
\usepackage{marginnote}
\usepackage[author=UTA]{pdfcomment}

\def\app#1#2{%
  \mathrel{%
    \setbox0=\hbox{$#1\sim$}%
    \setbox2=\hbox{%
      \rlap{\hbox{$#1\propto$}}%
      \lower1.1\ht0\box0}%
    \raise0.25\ht2\box2}%
}
\begin{document}

\title{Analytical model for polarization dependent light propagation in waveguide arrays and applications}

\author{S. Rojas-Rojas}
\affiliation{Center for Optics and Photonics and MSI-Nucleus on Advanced Optics, Universidad de Concepci\'{o}n, Casilla 160-C, Concepci\'{o}n, Chile}
\affiliation{Departamento de F\'{i}sica, Universidad de Concepci\'{o}n, Casilla 160-C, Concepci\'{o}n, Chile}

\author{L. Morales-Inostroza}
\affiliation{Center for Optics and Photonics and MSI-Nucleus on Advanced Optics, Universidad de Concepci\'{o}n, Casilla 160-C, Concepci\'{o}n, Chile}
\affiliation{Departamento de F\'isica, Facultad de Ciencias, Universidad de Chile, Santiago, Chile}

\author{U. Naether}
\affiliation{Instituto de Ciencia de Materiales de Arag\'on and Departamento de F\'{i}sica de la Materia Condensada, CSIC-Universidad de Zaragoza, 50009 Zaragoza, Spain}

\author{G. B. Xavier}
\affiliation{Center for Optics and Photonics and MSI-Nucleus on Advanced Optics, Universidad de Concepci\'{o}n, Casilla 160-C, Concepci\'{o}n, Chile}
\affiliation{Departamento de Ingenier\'{i}a El\'ectrica, Universidad de Concepci\'{o}n, Casilla 160-C, Concepci\'{o}n, Chile}

\author{S. Nolte}
\author{A. Szameit}
\affiliation{Institute of Applied Physics, Abbe Center of Photonics, Friedrich-Schiller-Universit\:at Jena, Max-Wien-Platz 1, Jena 07743, Germany}

\author{R. A. Vicencio}
\affiliation{Center for Optics and Photonics and MSI-Nucleus on Advanced Optics, Universidad de Concepci\'{o}n, Casilla 160-C, Concepci\'{o}n, Chile}
\affiliation{Departamento de F\'isica, Facultad de Ciencias, Universidad de Chile, Santiago, Chile}

\author{G. Lima}
\author{A. Delgado}
\affiliation{Center for Optics and Photonics and MSI-Nucleus on Advanced Optics, Universidad de Concepci\'{o}n, Casilla 160-C, Concepci\'{o}n, Chile}
\affiliation{Departamento de F\'{i}sica, Universidad de Concepci\'{o}n, Casilla 160-C, Concepci\'{o}n, Chile}

\begin{abstract}
We study the polarization properties of elliptical femtosecond-laser-written waveguides arrays. A new analytical model is presented to explain the asymmetry of the spatial transverse profiles of linearly polarized modes in these waveguides. This
asymmetry produces a polarization dependent coupling coefficient, between adjacent waveguides, which strongly affects the propagation of light in a lattice. Our analysis explains how this effect can be
exploited to tune the final intensity distribution of light propagated through the array, and links
the properties of a polarizing beam splitter in integrated optical circuits to the geometry of the waveguides.

\end{abstract}

\maketitle

\section{Introduction}

Integrated photonics is a very promising area of research for both classical and quantum phenomena due to highly controllable techniques for fabricating waveguide arrays~\cite{quantint,alex1}. The study of linearly polarized (LP) modes of the electromagnetic field in waveguides has mainly been addressed with the {\it weakly guiding approach}~\cite{wg,sy}. This model assumes that the refractive indices of core and cladding are nearly identical, simplifying the analysis by replacing the modes vectorial equation by a scalar equation. This leads to a degeneracy of both fundamental modes polarized along horizontal ($H$) and vertical ($V$) directions~\cite{sy}. Thereby, information related to the effect of polarization on the spatial transverse mode profiles is not considered. Nevertheless, differences between both LP modes are not negligible~\cite{sci}. In order to describe this phenomenon, a correction must be added to the solution of the the scalar equation. By considering the difference between dielectric constants for the core and the cladding as a perturbative parameter, a first order correction was formally proposed in~\cite{sl}. However, this approach does not properly predict the characteristic shape of each LP modal profile observed in~\cite{sci}.
An important application of photonic lattices is the analysis of the impact of disorder on light propagation, as they offer an ideal physical system to study the interplay of disorder and periodicity by means of simple table-top experiments. In fact, the first experimental demonstration of Anderson-localization was performed using optical lattices~\cite{naturesegev,1Dexp}. Disordered lattices exhibit a wealth of transport phenomena, such as disorder-induced edge states \cite{edgedisorder}, disorder-enhanced transport \cite{levi,stutz}, and the interplay between nonlinearity and disorder~\cite{oeus}.

Here, we present a new approach for modelling the polarization dependence of electric field profiles of the tranverse modes in elliptical waveguide arrays and, consequently, of coupling constants between neighbor waveguides. Furthermore, we study the interplay between this latter effect and the inclusion of disorder. We experimentally corroborate our findings with femtosecond-laser-written elliptical waveguide arrays in silica substrates~\cite{fs} with off-diagonal disorder, by controlling the input polarization of an initially very localized excitation. By varying the initial polarization vector we are able to {\it tune} the localization volume of the light propagated through the array. Finally, our model can be applied to design  polarizing beam splitters (PBS) in integrated photonic circuits. In a PBS, the coupling coefficients define the transmittance for different polarizations. The model links these transmittances  to the geometrical properties of the waveguides. This result has multiple applications in the area of quantum information since a PBS corresponds to a CNOT gate \cite{quantel,qucomp}.

\section{Polarization dependence of spatial mode profiles}\label{sc:model}

\begin{figure}[t]
 \centering
 \includegraphics[width=.44\textwidth]{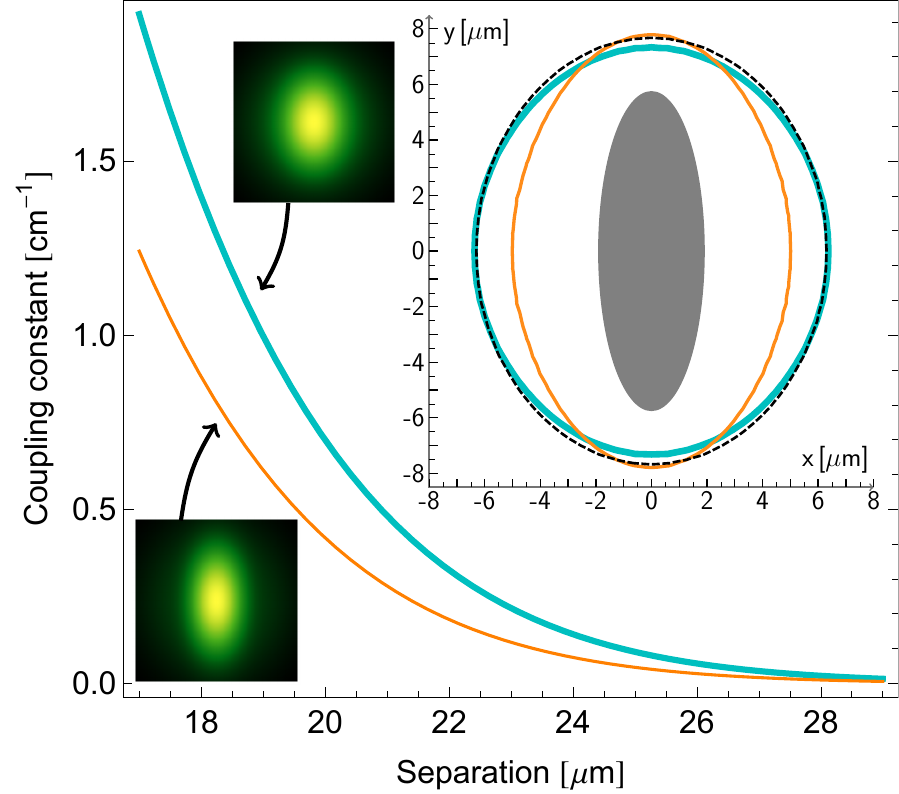}
 \caption{The coupling constant between two adjacent waveguides is shown for $H$ ($V$) polarization in blue thick (orange thin) line. We also show in the insets the amplitude of the electric field for the obtained modes, which were used to compute the coupling constants, and the corresponding contours for the LP modes. Black dashed line represents the contour of the modal profile of the zero-order
 solution $\tilde e$ of the scalar equation. Grey region is the transverse section of the waveguide core.
 }\label{fg:cpfs}
\end{figure}

The simplest approximation to the LP modes is obtained with the equation 
\begin{equation}
\left\lbrace {\nabla}_t^2+ \left[k^2n^2(x,y)-\beta^2\right]\right\rbrace e(x,y)=0\,, 
\end{equation}
where $e(x,y)$ is a linearly polarized electric field with its corresponding propagation constant $\beta$, ${\nabla}_t^2$ is the transversal Laplacian operator, and $k=2\pi/\lambda$ 
is the wave number. $n(x,y)$ is the refractive index, which we consider to have a specific value for the core and the cladding. We solve the scalar equation by using a finite-element method and label the solution as the {\it zero-order} approximation $\tilde e$ for both, $H$ and $V$, fundamental modes \cite{fem}. 
Then, from Maxwell Equations the longitudinal magnetic component related to the approximated field ${\bf \tilde e}$ is obtained:
\begin{equation}\label{eq:hz}
\begin{split}
 h_z&=-i\sqrt{\frac{\epsilon_0}{\mu_0}}\frac{1}{k}\hat{\bf z}\cdot {\bm \nabla}_t\times{\bf e}_t\\
 &\simeq -i\Delta^{1/2}\sqrt{\frac{2\epsilon_0}{\mu_0}}\frac{\rho n_{\rm core}}{v}\hat{\bf z}\cdot {\bm \nabla}_t\times{\bf \tilde e}\equiv\Delta^{1/2}h_z^{(1/2)}\,,
\end{split}
\end{equation} 
with the {\it waveguide parameter} $v$ defined by $k\rho\sqrt{n_{\rm core}^2-n_{\rm cladd}^2}$, and $\Delta$ being the relative difference between the dielectric constants in the core and the cladding: $(n_{\rm core}^2-n_{\rm cladd}^2)/2n_{\rm core}^2$. $\rho$ must be a scale length  characteristic of the waveguide, so we set it as the mean radius of the tranverse section. Notation $h_z^{(1/2)}$ is used just to denote the whole factor that goes along with $\Delta^{1/2}$. The curl in the definition of $h_z$ has a clearly different effect on each polarization. By using Maxwell Equations again, and assuming that the fields are TE, we obtain a transverse correction of order $\Delta$, proportional to $\hat{z}\times {\bf \nabla}h_z$, which also depends on polarization. Thereby, we obtain a different profile for each mode $e_H$ and $e_V$. The shape of the transverse electric field profile for each LP mode is shown in the insets of Fig.~\ref{fg:cpfs}, where we have taken $x$ ($y$) as the direction of $H$ ($V$) polarization (the relevant parameters have been chosen to coincide with those of our experiment). Clearly, each polarization mode has a characteristic and different profile. In our model, propagating modes of elliptical waveguides are {\it hybrid}, as they have both electric and magnetic components~\cite{agrawal,yeh}.

%

We are interested in the propagation of light on an array of waveguides along the $x$ direction. We resort to the coupled mode theory,
where light couples between nearby waveguides at a rate given by the {\it coupling constant}~\cite{sl}
\begin{equation}\label{eq:c}
 C=-\frac{k}{2n_{\rm core}}\frac{\int dxdy\, \left[n(x,y)-n_{\rm core}\right] e(x,y)e(x-s,y)}{\int dxdy\,e(x,y)^2}\ ,
\end{equation}
where $e$ is the transversal electric field corresponding to a normal mode (either $e_H$ or $e_V$). $n(x,y)$ represents the refractive index pattern, with the corresponding core and cladding structure. $s$ corresponds to the separation between neighbor waveguides. Here, only nearest-neighbour coupling is considered. From this definition, and our previous results for the polarization dependent modal profiles, we obtain two different coupling constants $C_H$ and $C_V$, for each polarization. Fig.~\ref{fg:cpfs} shows the value of these constants as a function of the distance between two neighboring waveguides. From this figure, we observe clearly how different polarizations experience different coupling coefficients, being this a stronger (weaker) effect for smaller (larger) separation distances. By extending the coupled mode approach to an array of waveguides~\cite{led}, we arrive to the set of equations that governs the evolution of the light amplitude $u_n(z)$, along the propagation direction $z$, at the $n$-th guide of the array
\begin{equation}\label{eq:dnls}
 -i\frac{d }{dz}u_{n}^{\sigma}(z)=C_{n,n+1}^\sigma u_{n+1}(z)+C_{n,n-1}^\sigma u_{n-1}(z)\,,
\end{equation}
for $\sigma=H,V$. The field in each site is given by $u_n (z)\cdot e(x-x_n,y)$, where $x_n$ corresponds to the central position at the $n$-th site. $C_{n,n'}^\sigma$ is the coupling constant between sites $n$ and $n'$, for polarization $\sigma$.

\section{Experimental results}

\begin{figure}[t]
 \centering
 \includegraphics[width=.48\textwidth]{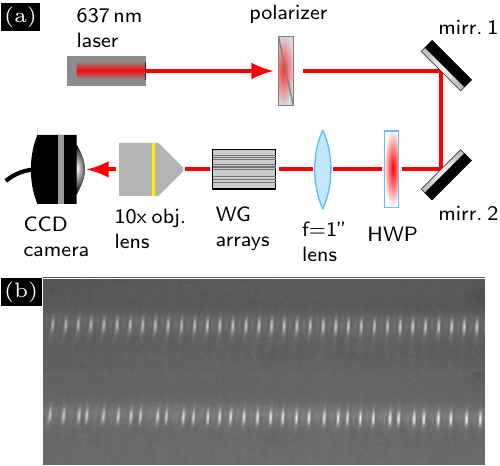}
  \caption{Experimental setup. (a) A laser beam, with defined polarization, is focused into a sample with several waveguide arrays. The output
  profile is imaged with a $10\times$ objective onto a CCD camera. (b) Microscope image of the waveguides facet for an ordered (top) and a disordered (bottom) lattice, illuminated with a wide beam of white light.}\label{fg:setup}
\end{figure}

\begin{figure}[b!]
 \centering
 \includegraphics[width=.48\textwidth]{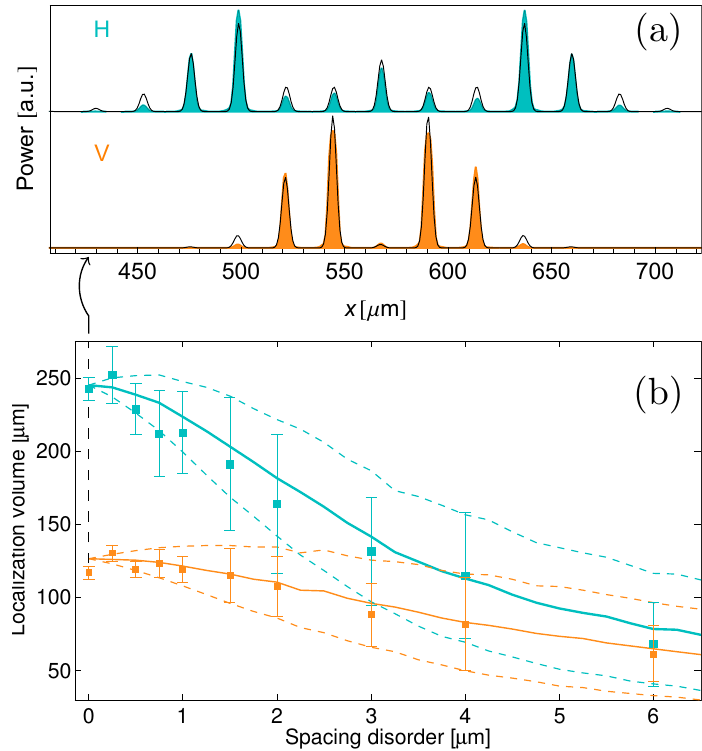}
  \caption{(a) Output profile obtained for an ordered array: filled area correspond to experimental data, while black lines
  to theoretical results of Eq.~(\ref{eq:dnls}). (b) Average output localization volume versus disorder strength: lines correspond to
  theory and squares to the experimental data. Dashed lines (bars) show the standard deviation of theoretical (experimental) results at each point. Blue (orange) color
  corresponds to $H$ ($V$) polarization.  }\label{fg:cons}
\end{figure}

First, we study experimentally the propagation of light in an ordered elliptical waveguide array fabricated in fused silica by the femtosecond laser writing technique~\cite{fs}. The array consists of $71$ equally spaced waveguides, with a separation of $23\ \mu $m, and a total propagation length of $10$ cm. The waveguides have an elliptic profile with major and minor axis of 12 $\mu m$ and 4 $\mu m$, respectively. We excite the array by focusing a $637$ nm CW laser beam into a single waveguide (single-site excitation), and record the output intensity with a CCD camera [see Fig.~\ref{fg:setup}].  As expected, discrete diffraction~\cite{led,alex1} was observed for both $H$ and $V$ polarizations [see Fig.~\ref{fg:cons} (a)]. If the input waveguide is $n=0$, the field amplitude in each waveguide, at propagation distance $z$, can be expressed as~\cite{led}: $u_n^\sigma(z)\propto i^n J_n(2C^{\sigma} z)$, where $J_n$ is the $n$-th order Bessel function. We experimentally measure the output profile at $z_f=10$ cm, and determine $C^{\sigma}$. In order to compare our simulations with the experimental results, we assume a gaussian distribution in every waveguide output. In this way, we get a continuous field distribution $U^{\sigma}(x,z_f)$. We find the coupling constant to be $(0.223\pm 0.001)\,{\rm cm}^{-1}$ for $H$ polarization, and $(0.112\pm0.001)\,{\rm cm}^{-1}$ for $V$ polarization in close agreement to the theoretical predictions (see curves in Fig. \ref{fg:cpfs} at separation 23\,$\mu$m). So, we find a factor close to $2$ between the determined coupling constants of both polarizations. The errors of the constants were obtained by minimizing the squared two-norm of the residuals between analytical and experimental results. With the coupling constants values, we estimate the contrast $\Delta n$ between the core and the cladding refractive indices. We determine that our waveguides have a refractive index contrast of $\Delta n=9.37\times 10^{-4}$, which is in agreement with the reported values in literature \cite{alex1}. This value was obtained under the assumption of a step-index profile. Real waveguides exhibit a continuous but sharp profile, being our $\Delta n$ a good approximation to the contrast between the core center and the cladding.
%

\begin{figure}[!t]
 \centering
 \includegraphics[width=.48\textwidth]{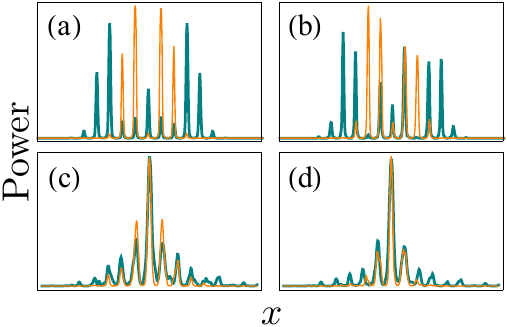}
 \caption{Experimental averaged output for $H$ (blue line) and $V$ (orange line) polarized light. (a) Ordered case. (b) Disorder parameter $\epsilon=0.75\,\mu$\,m. (c)  $\epsilon=3\,\mu$m. Notice the exponential decay in both sides of the profile. (d) $\epsilon=6\,\mu$m. In these cases, the decay is more clear and similar for both polarizations.}\label{fg:pfdes}
\end{figure}

We now consider a more complex case, where the coupling constant depends on the polarization $\sigma$, as well as on the particular pair of neighboring lattice sites, say $n$ and $n'$. If the value of the coupling constant $C_{n,n'}^\sigma$ is varied at random across the array, this introduces disorder in the system. Experimentally, this is achieved by randomly varying the separation between each pair of waveguides during the fabrication of the array, in a range $\overline s \pm \epsilon$, where $\overline s$ is the mean separation and the $\epsilon$ parameter is called the {\it spacing disorder}. We study the effect of disorder on the propagation of a single-site initial excitation of the form: $u_n^{\sigma}(0)=\delta_{n,n_0}$, with $n_0$ the input position. First, we solve numerically the respective set of equations~\eqref{eq:dnls} up to a distance $z_f=10$\,cm. For each degree of disorder, we average the resulting profile by simulating $1000$ different realizations, i.e. different distributions of separations between waveguides. To characterize the output profiles, we choose the localization volume~\cite{fk}, defined as $V_c\equiv(\sqrt{12 m_2}+1)\,[\mu m]$, where $m_2$ corresponds to the profile's second moment, given by
\begin{equation}\label{eq:m2}
 m_2\equiv\frac{\int (x-\overline x)^2|U^{\sigma}(x,z_f)|^2\,dx}{\int |U^{\sigma}(x,z_f)|^2\,dx}\,,
\end{equation}
with $\overline x\equiv\int x|U(x)|^2\,dx$. The parameter $V_c$ provides an estimation of the distance between the exponential tails of the profile, and it can be computed directly from the numerical and the experimental data. The theoretical result, presented by full lines in Fig.~\ref{fg:cons}(b), show a decaying tendency of the localization volume for an increasing degree of disorder, for both polarizations. This agrees with the known results
on disordered lattices: an extinction of diffusion for an increasing disorder~\cite{naturesegev,1Dexp,oeus}. The polarization effect is most prominent in the ordered lattice due to the different spreading rates of the ballistic lobes.

As said before, experimental disorder is introduced by randomly varying the separation between waveguides. We study a set of nine disordered waveguides arrays, where the spacing between guides lies in the range $23\pm\epsilon\ \mu$m, with $\epsilon=(0.25,0.5,0.75,1,1.5,2,3,4,6)\ \mu$m [as an example, see Fig.~\ref{fg:setup}(b)]. Thus, we analyze the effect of weak, intermediate and strong disorder. Following the usual method (cfr. Ref. \cite{1Dexp}), we used $40$ different input waveguides in each array, in order to have significant statistics. Symbols in Fig.~\ref{fg:cons}(b) show the experimentally averaged localization volume of the output profiles. We observe how the initial large difference for $V_c$, for the $H$ and $V$ polarized light, decrease due to Anderson localization in disordered lattices. The exponential decay of the output profiles far from the input position, characteristic of this phenomenon
, can be appreciated in Fig. \ref{fg:pfdes}. 
Diagrams of $V_c$ vs disorder exhibit very good agreement between theory and the experimental results for both $H$ and $V$ polarization, with a normalized squared euclidian distance \cite{note} of $0.006$ and $0.009$ respectively (null distance corresponds to perfect match and 1 implies no correlation at all).

\begin{figure}[t!]
 \includegraphics[width=.45\textwidth]{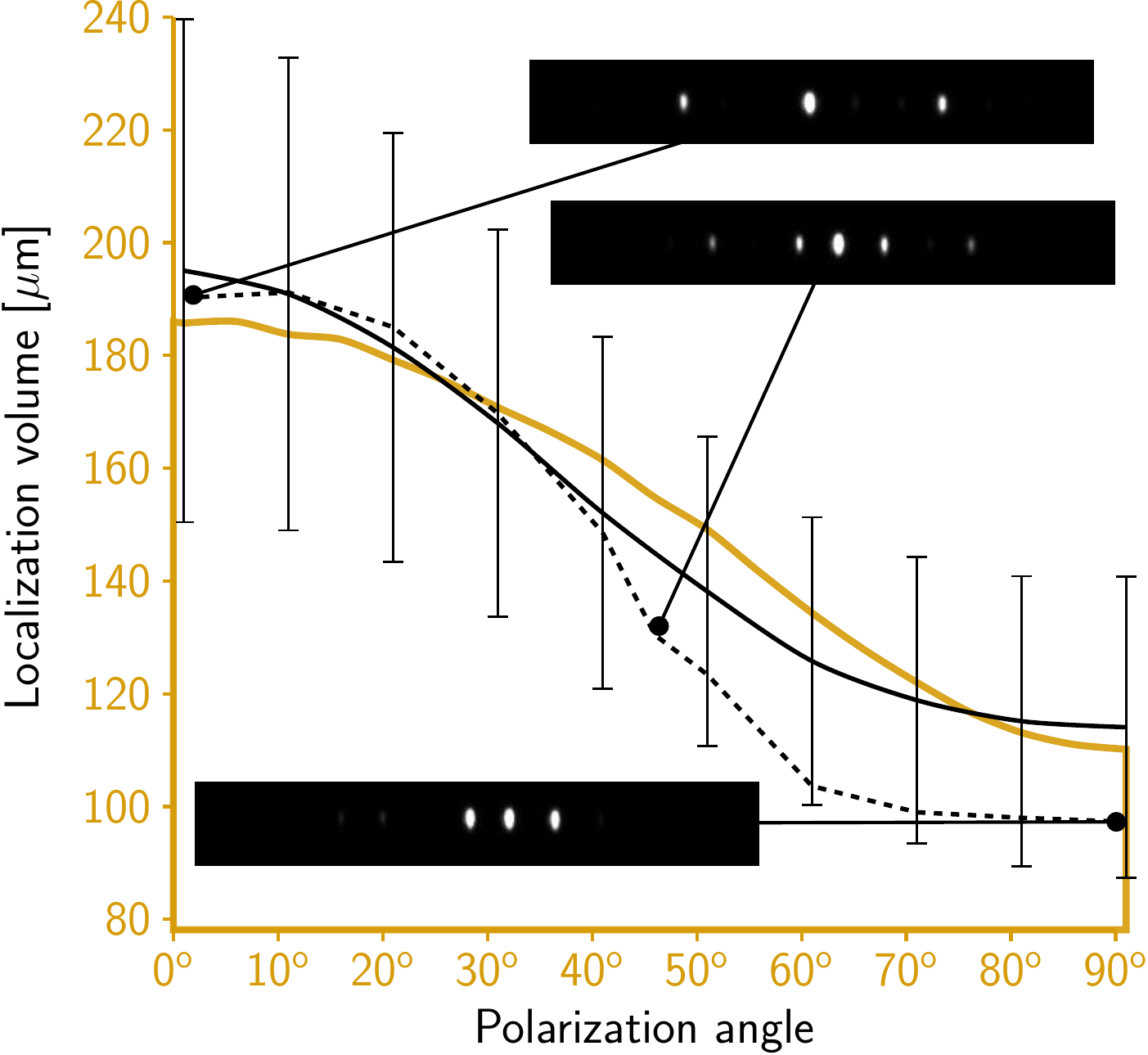}
 \caption{Localization volume versus polarization angle for $\epsilon=2$ $\mu$m. Full black line describes the averaged experimental results. Dashed line and insets show the data obtained for one realization. Thick line shows averaged theoretical results.
 }\label{fg:tuning}
\end{figure}

\section{Tuning the localization volume}

The clear difference observed in the propagation of $H$ and $V$ polarized beams suggests the possibility to control the localization volume by tuning the initial polarization vector. We illuminated a single waveguide in different samples and varied the polarization angle from 0\textordmasculine ($H$) to 90\textordmasculine ($V$). In average, we observe a smooth transition of the localization volume for weak and intermediate disorder. See an example in Fig.~\ref{fg:tuning}. For $H$ polarized light (0\textordmasculine), the state corresponds to the excitation of three separated waveguides, one at the input position (center) and two waveguides $3$ sites away from the center [see inset-top in Fig.~\ref{fg:tuning}].  As the polarization flips, a smooth attenuation of the amplitude on these two surrounding waveguides occurs, as light begins to keep focused around the center. For example, for $\theta=45$\textordmasculine, 
additional excitation of two sites next to the center guide is observed [see inset-middle in Fig.~\ref{fg:tuning}]. This intermediate state corresponds to a linear combination of $H$ and $V$ states
. For $V$ polarization ($\theta=90$\textordmasculine), the state corresponds to essentially three equally excited neighboring waveguides centered at the input position [see inset-bottom in Fig.~\ref{fg:tuning}]. Therefore, we were able to tune the localization volume and observe a decrease to a half of the initial distribution volume.

\section{Design of a polarizing beam splitter}

\begin{figure}[!b]
 \includegraphics[width=.42\textwidth]{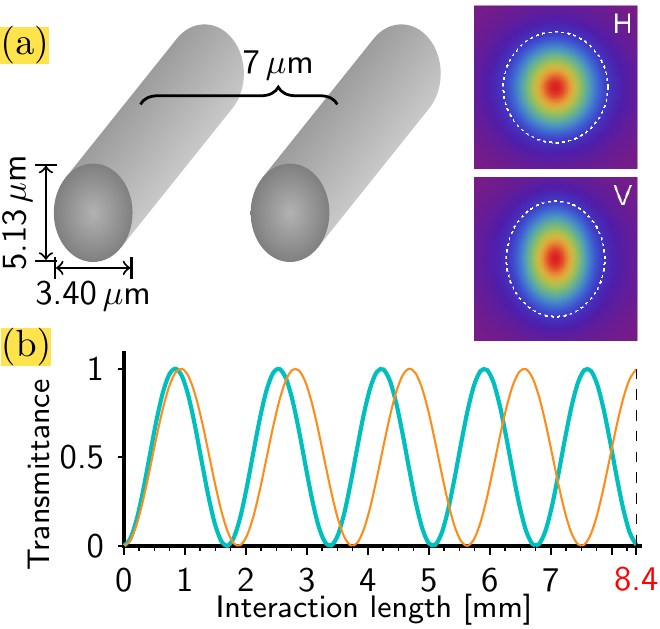}
 \caption{(a) Two-waveguides configuration to produce a compact, balanced and deterministic PBS. Density plots of the transverse electric fields profiles are shown. (b) Transmittance of $H$ (thick blue line) and $V$ (thin orange line) polarized light.}\label{fg:pbs}
\end{figure}

In order to further test our model, we now consider
a system composed by only two waveguides, i.e., a linear dimer coupler. In this case, when light is injected to a single guide, the energy periodically transfers from one waveguide to the other at a rate given by the coupling constant. That is, the transmittance will depend on the coupling constant. With the appropriate values of the coupling constants for each linear polarization, $H$ and $V$, it is possible that after a fixed propagation length, light with a certain polarization will propagate in one waveguide and the light with the opposite polarization in the other waveguide (See Fig. \ref{fg:pbs}). This idea has been used in previous works to construct polarizing beam splitters (PBS) \cite{Fer1,fabPBS}. 
Now we show that with the treatment of Section \ref{sc:model}, it is possible to find a suitable geometrical configuration of the waveguides to obtain a polarizing beam splitter with high splitting ratio and an interaction length in the order of milimeters.

To design a PBS we study the transmittance $T_\sigma$ of light from one waveguide to the other. This is a function of the coupling constant $C_\sigma$ and the interaction length $L$ by the expression
\begin{equation}\label{eq:T}
 T_\sigma=\sin^2(C_\sigma L)\,,
\end{equation}
where $\sigma=H,V$. In setups where waveguides are brought together adiabatically through a bending region, a phase term must be added in the argument of the sine function. In order to have complete separation of the $H$ and $V$ components of a light beam, the following condition must be fulfilled:
\begin{equation}\label{eq:raz}
 \frac{C_V}{C_H}=\frac{m}{2n}\,,
\end{equation}
with $m$ odd and $n$ integer. The corresponding interaction length is then given by $L_{\rm split}=\pi n/C_H$. Thus, we look for values of the coupling constants which minimize $L_{\rm split}$. In general, the ratio $C_V/C_H$ has to be approximated to the rational expression Eq. \eqref{eq:raz} by an error term $\delta$. From Eq. \eqref{eq:T} it can be seen that up to first order, this error term will propagate linearly to transmittance. We impose that $\delta=5\times 10^{-4}$  and varied the ellipticity and separation between the waveguides.  We found that with a separation between waveguides of 7\,$\mu$m and for semi-axes of 1.70\,$\mu$m  and 2.56\,$\mu$m (See Fig. \ref{fg:pbs}), a balanced PBS can be fabricated with coupling constants $C_H=1.8609\,{\rm mm}^{-1}$ and $C_V=1.6754\,{\rm mm}^{-1}$, which lead to an interaction length of only 8.4\,mm. This is consistent with the results reported in refs. \cite{Fer1,fabPBS}.

\section{Conclusion}

We have presented a new model to describe the effects of polarization onto the spatial modes of light propagating in arrays of waveguides. These theoretical findings are in agreement with our experimental results obtained using fs-written elliptical waveguide arrays. It was also possible to determine the refractive index contrast between the core and the cladding of the waveguides. The experimentally obtained parameters were used to analyze the value of the localization volume depending on the degree of disorder. Consistently, the localization volume for a fixed disorder degree is smaller for vertically polarized light, indicating that this type of polarization produces a stronger localization although this effect tends to vanish, as expected, in the strong Anderson localization regime. We used the dependence of the coupling constant on the polarization to tune the localization volume. A potential extension of this work is the study of two-dimensional (2D) arrays. For 2D systems, we do not expect to experimentally observe similar results. As a larger disorder is required to observe localization in 2D lattices \cite{oeus}, the whole picture will be scaled up and the different polarizations will essentially show the same. The transition to localization in 1D is more abrupt and, therefore, the different polarizations really experience different dynamics.

Finally, our model, which provides a link between the geometry of the waveguides, the polarization of light and the coupling constants, allowed us to find a feasible experimental configuration to produce a compact and balanced PBS that is useful for many quantum information tasks. We  remark that although currently there exist polarizing beam splitter cubes of about $5\,$mm$^3$, there are great advantages in the use of waveguides to implement this device \cite{qw}.

\acknowledgments

Authors acknowledge funding from FONDECYT Grants 1110142, 1120067, Programa ICM P10-030-F, Programa de Financiamiento Basal de CONICYT (FB0824/2008), the
Spanish government project FIS 2011-25167, the German Ministry of Education and Research (Center for Innovation Competence program, grant 03Z1HN31),
Thuringian Ministry for Education, Science and Culture (Research group Spacetime, grant no. 11027-514), the Singapore National Research Foundation
and Ministry of Education (partly through the Academic Research Fund Tier 3 MOE2012-T3-1-009) and the German-Israeli Foundation for Scientific
Research and Development (grant 1157-127.14/2011).



\begin{thebibliography}{99}

\bibitem{quantint} J. L. O'Brien, A. Furusawa and J. Vuckovic, 
Nature Photonics \textbf{3}, 687 - 695 (2009).

\bibitem{alex1} A. Szameit and S. Nolte, 
J. Phys. B: At. Mol. Opt. Phys. {\bf 43}, 163001 (2010).

\bibitem{wg} D. Gloge, 
Appl. Opt. {\bf 10}, 2252 (1971).

\bibitem{sy} A. W. Snyder and W. R. Young, 
J. Opt. Soc. Am {\bf 68}, 297 (1978).

\bibitem{sci} L. Sansoni, F. Sciarrino, G. Vallone, P. Mataloni, A. Crespi, R. Ramponi and R. Osellame, 
Phys. Rev. Lett. {\bf 108}, 010502 (2012).

\bibitem{sl} A. W. Snyder and J. D. Love, {\it Optical Waveguide Theory}, Chapman and Hall, New York (1983).

\bibitem{naturesegev} T. Schwartz, G. Bartal, S. Fishman and M. Segev, 
Nature {\bf 446}, 52-55 (2007).

\bibitem{1Dexp} Y. Lahini, A. Avidan, F. Pozzi, M. Sorel, R. Morandotti, D. N. Christodoulides, and
Y. Silberberg, 
\prl {\bf 100}, 013906 (2008).

\bibitem{edgedisorder} A. Szameit, Y. V. Kartashov, P. Zeil, F. Dreisow, M. Heinrich, R. Keil, S. Nolte, A. T\"{u}nnermann, V. A. Vysloukh, and L.
Torner, 
Opt. Lett. \textbf{35}, 1172 - 1174 (2010).

\bibitem{levi} L. Levi, M. C. Rechtsman, B. Freedman, T. Schwartz, O. Manela, and M. Segev, 
Science \textbf{332}, 1541 - 1544 (2011).

\bibitem{stutz} S. St\"{u}tzer, T. Kottos, A. T\"{u}nnermann, S. Nolte, D. N. Christodoulides, and A. Szameit, 
Opt. Lett. \textbf{38}, 4675 - 4678 (2013).

\bibitem{oeus} U. Naether, S. Rojas-Rojas, A. J. Mart\'inez, S. St\"utzer, A. T\"unnermann, S. Nolte, M. I. Molina, R. A. Vicencio, and A. Szameit, 
Opt. Express {\bf 21}, 927 (2013).

\bibitem{fs} A. Szameit, F. Dreisow, T. Pertsch, S. Nolte and A. T\"unnermann, 
Opt. Exp. {\bf 15}, 1579 - 1587 (2007).

\bibitem{quantel} C. H. Bennett, G. Brassard, C. Cr\'epeau, R. Jozsa, A. Peres, and W. K. Wootters, Phys. Rev. Lett. {\bf 70}, 1895 (1993).

\bibitem{qucomp} H. J. Briegel and R. Raussendorf, Phys. Rev. Lett. {\bf 86}, 910 (2001); R. Raussendorf and H. J. Briegel, Phys. Rev. Lett. {\bf 86}, 5188 (2001).

\bibitem{fem} C. Yeh, K. Ha, S. B. Dong and W. P. Brown, 
Appl. Opt. {\bf 18}, 1490 (1979).

\bibitem{agrawal} G. P. Agrawal, {\it Nonlinear Fiber Optics} 3rd ed., Academic Press (2001).

\bibitem{yeh} C. Yeh, 
J. Appl. Phys. {\bf 33}, 3325 (1962).

\bibitem{led} F. Lederer, G. I. Stegeman, D. N. Christodoulides, G. Assanto, M. Segev, and Y. Silberberg, 
Phys. Rep. \textbf{463}, 1 (2008).

\bibitem{fk} D. O. Krimer and S. Flach, Phys. Rev. E {\bf 82}, 046221 (2010).

\bibitem{note} The {\it normalized squared euclidian distance} between two sets of data $x_i$ and $y_i$, whose respective means are $\overline x$ and $\overline y$, is given by $$\frac{1}{2}\frac{\sum_i\left|(x_i-\overline x)-(y_i-\overline y)\right|^2}{\sum_i\left|x_i-\overline x\right|^2+\sum_i\left|y_i-\overline y\right|^2}\,.$$

\bibitem{Fer1} L. A. Fernandes, J. R. Grenier, P. R. Herman, J. S. Aitchison, and P. V. S. Marques, Opt. Exp. {\bf 19} 11992 (2011).

\bibitem{fabPBS} A. Crespi, R. Ramponi, R. Osellame, L. Sansoni, I. Bongioanni, F. Sciarrino, G. Vallone, and P. Mataloni, Nat. Commun. {\bf 1570}, 566 (2011).




\bibitem{qw} A. Peruzzo, M. Lobino, J. C. F. Matthews, N. Matsuda, A. Politi, K. Poulios, X-Q Zhou, Y. Lahini, N. Ismail, K. W\"orhoff, Y. Bromberg, Y. Silberberg, M. G. Thompson, and J. L. OBrien, Science {\bf 329}, 1500 (2010).


\end{thebibliography}
\end{document}